\documentclass[aps,pre,epsfig,floats,twocolumn,superscriptaddress,amssymb,amsmath,floatfix,showpacs]{revtex4}
\usepackage{graphicx}
\begin{document}

\title{A Simplest Swimmer at Low Reynolds Number: Three Linked Spheres}

\author{Ali Najafi}
\affiliation{Institute for Advanced Studies in Basic Sciences,
Zanjan 45195-159, Iran}

\affiliation{Max-Planck Institut  f\"{u}r Physik  Komplexer
Systeme, N\"{o}thnitzerstrasse 38, 01187 Dresden, Germany}

\author{Ramin Golestanian}
\affiliation{Institute for Advanced Studies in Basic Sciences,
Zanjan 45195-159, Iran}

\date{\today}

\begin{abstract}
We propose a very simple one-dimensional swimmer consisting of
three spheres that are linked by rigid rods whose lengths can
change between two values. With a periodic motion in a
non-reciprocal fashion, which breaks the time-reversal symmetry as
well as the translational symmetry, we show that the model device
can swim at low Reynolds number. This model system could be used
in constructing molecular-size machines.
\end{abstract}

\maketitle
The usual swimming mechanism for a human being in water involves
obtaining a forward momentum from the surrounding fluid due to
some periodic body motion. The fact that the displacement gained
in the first half period of the cyclic motion is not canceled out
by that of the second half period is known to be predominantly
because of the inertial effects \cite{childress}. Such a
mechanism, however, does not work in the microscopic world of
biological objects (such as bacteria), where the effects of
inertia are not important and the viscous effects dominate. This
case is characterized by very low Reynolds number---the
dimensionless quantity that measures the ratio between the
inertial term and the viscous term in the hydrodynamical equation
of motion \cite{lrnh}.

Most microscopic biological objects can swim very well with
velocities of the order of $1 \; \mu {\rm m}/{\rm s}$, which for
such micron-sized animals swimming in water, yields Reynolds
numbers of the order of $10^{-4}$. In his pioneering work, Purcell
showed that animals like scallop that are equipped with a single
hinge cannot swim, using a simple opening and closing procedure
\cite{purcell1,purcell2}. The reason is simple: since the motion
is reversible, after finishing a cycle the animal will end up
being where it initially was. He proposed that a non-reciprocal
motion, which breaks the time-reversal symmetry, is needed to
produce a net displacement. This will help the animal to propel
itself, during each cycle, along some direction that is preferred
by the symmetry of the system and the motion \cite{taylor}.
Despite the simplicity of the geometry that he suggested---three
rigid rods connected with two hinges---quantitative analysis of
the ``Purcell swimmer'' had not been performed until very
recently, due to the complexity of the Stokes equation in the
specific geometry \cite{howard1}. We note that there are other
related swimming mechanisms at the mesoscopic scale in the context
of molecular machines \cite{porto}.

Here we use Purcell's original idea and introduce a very simple
and experimentally accessible model system that can swim using
proposed periodic internal motions. The swimmer consists of three
hard spheres that are linked through two arms, and has the
advantage that the details of the hydrodynamic interactions, as
well as the swimming velocity and direction, can be worked out
with great ease, as compared to the case of the Purcell swimmer.

\begin{figure}
\includegraphics[width=.7\columnwidth]{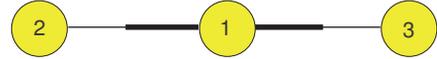}
\caption{Three linked spheres connected by two arms of negligible
thickness.} \label{swimmer}
\end{figure}

\begin{figure}
\includegraphics[width=.7\columnwidth]{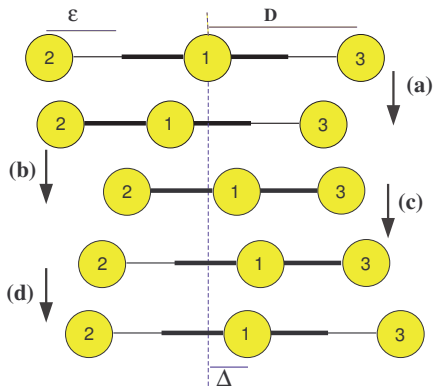}
\caption{Complete cycle of the proposed non-reciprocal motion of
the swimmer, which is composed of 4 consecutive time-reversal
breaking stages: (a) the left arm decreases its length with the
constant relative velocity $W$, (b) the right arm decreases its
length with the same velocity, (c) the left arm opens up to its
original length, and finally, (d) the right arm elongates to its
original size. By completing the cycle the whole system is
displaced to the right side by an amount $\Delta$.}
\label{nonrecipcycle}
\end{figure}

The model swimmer that we are proposing is shown in Fig.
\ref{swimmer}, and consists of three spheres with radius $R$ that
are connected by rigid slender arms aligned along the $x$
direction. The spheres are assumed to be floating in a highly
viscous fluid with viscosity $\mu$. There are two internal engines
on the middle sphere (sphere number 1), which act as internal
active elements responsible for making a non-reciprocal motion
that is needed to propel the whole system. We consider the initial
state of the system such that the spheres number 2 and 3 are in
equal distance $D$ from the middle sphere. We divide a complete
cycle of the non-reciprocal motion into four parts as below (see
Fig. \ref{nonrecipcycle}):

\noindent(a) In the first step of the motion, the right arm has
fixed length, and the length of the left arm is decreased with a
constant relative velocity $W$, using one of the internal engines
in the middle sphere. We denote the relative displacement of the
spheres 1 and 2 in this stage by $\epsilon$.

\noindent(b) As the second step, the left arm is fixed and the
right arm decreases its length with the same constant relative
velocity $W$ as before. The relative displacement of the spheres 1
and 3 is again $\epsilon$, like the previous stage.

\noindent(c) During this step, while the right arm is kept fixed,
the left arm increase its length with the same relative velocity
$W$ to reach its original length $D$.

\noindent(d) Finally, in the last step the left arm is kept fixed
and the right arm elongates to its original length with the same
constant velocity $W$. The system is now in its original internal
configuration.

As can be seen from Fig. \ref{nonrecipcycle}, the above 4-stage
cycle is not invariant under time reversal, and we can thus expect
a net translation upon completing a full cycle. To obtain a net
translational motion, the above cycle can be repeated
continuously.

The general equation that describe the hydrodynamics of low
Reynolds number flow is the Stokes equation for the velocity field
${\bf u}$, subject to the incompressibility condition:
\begin{eqnarray}
&&\mu\nabla^{2}{\bf u}-\nabla p=0,\label{stokes} \\
&&\nabla \cdot{\bf u}=0, \label{compresibility}
\end{eqnarray}
where $p$ represents the pressure field in the medium.

Assuming that the spheres are moving inside the fluid with
velocity vectors ${\bf V}_{i}$, with the index $i$ denoting the
labels of the spheres, the description of the system involves
solving the above-mentioned equations with zero velocity boundary
condition at infinity and no-slip boundary conditions on the
spheres, which implies
\begin{equation}
{\bf u}|_{{\bf r}~{\rm on~the~}i{\rm 'th~sphere}}={\bf
V}_{i}.\label{BC}
\end{equation}
The variables that are necessary to determine the dynamics of the
spheres are their velocities ${\bf V}_{i}$ and the forces ${\bf
F}_{i}$ acting on them. By solving the above governing equations,
we will be able to obtain the fluid velocity in the medium, and
hence, the corresponding stress tensor that will give us the
required forces on the spheres. It is a simple observation that
because of the linearity of the governing equations, and the
linearity of the stress tensor with respect to the velocity field,
one can generally expect a relation of the form:
\begin{equation}
{\bf V}_{i}=\sum_{j=1}^{3}{\cal H}_{ij} \cdot {\bf F}_{j},
\label{forcevelocity}
\end{equation}
where the symmetric Oseen tensor ${\cal H}_{ij}$ depends on
viscosity, the geometry of the bodies immersed in it (in our case
the spheres), and their relative orientations. Including the
condition that there are no external forces such as gravity, the
system of spheres should be force-free:
\begin{equation}
\sum_{j=1}^{3}{\bf F}_{j}=0.\label{forcefree}
\end{equation}

\begin{figure}
\includegraphics[width=.7\columnwidth]{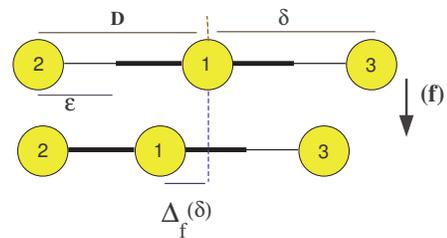}
\caption{An auxiliary (fictitious) movement in which the right arm
has a constant length $\delta$ while the left arm changes its
length from $D$ to $D-\epsilon$. During this movement the middle
sphere will be displaced by an amount $\Delta_f(\delta)$.}
\label{auxiliary}
\end{figure}

Since we are only interested in the dynamics of the spheres, we
can equivalently solve the set of Eqs. (\ref{forcevelocity}) and
(\ref{forcefree}), instead of Eqs. (\ref{stokes}) and
(\ref{compresibility}). To solve these equations we need to know
the form of the Oseen's tensor. Let us consider a coordinate
system in which the position vector of the $i$'th sphere is ${\bf
x}_{i}$ and the separation between the $i$'th and the $j$'th
spheres will be ${\bf x}_{ij}\equiv{\bf x}_{i}-{\bf x}_{j}$, with
a unit vector $\hat{{\bf n}}$ in this direction. General symmetry
considerations will allow us to write the hydrodynamic interaction
tensor in the following form \cite{batchelor1,batchelor2}:
\begin{equation}
{\cal H}_{ij}=\frac{1}{6\pi\mu R}[A_{ij}(\lambda)\hat{{\bf
n}}\hat{{\bf n}}+B_{ij}(\lambda)({\bf I}-\hat{{\bf n}}\hat{{\bf
n}})].
\end{equation}
where we use the dimensionless quantity $\lambda={R}/{{x}_{ij}}$.
Assuming that the separations between the spheres are sufficiently
larger than their sizes, we can write a perturbation expansion for
the symmetric coefficients $A_{ij}$ and $B_{ij}$ in powers of
$\lambda$, which reads
\begin{equation}
\begin{array}{l}
A_{ij}=\left\{
  \begin{array}{l}
    1+O(\lambda^{4}),\;\;\;\;\;\;\;\;\;\;\; i=j\\
    \frac{3}{2}\lambda_{ij}+O(\lambda^{3}),\;\;\;\;\;\; i \neq j\\
  \end{array}
                   \right.
\end{array}
\end{equation}
and
\begin{equation}
\begin{array}{l}
B_{ij}=\left\{
  \begin{array}{l}
    1+O(\lambda^{4}),\;\;\;\;\;\;\;\;\;\;\; i=j\\
    \frac{3}{4}\lambda_{ij}+O(\lambda^{3}),\;\;\;\;\;\; i \neq j\\
  \end{array}
                   \right.
\end{array}
\end{equation}
to the leading order. Note that for simplicity we are considering
only the translational motion for the spheres, although the
effects of rotational motion can be taken into account in a
similar way. Moreover, extra simplification comes from the
one-dimensional nature of the motion, such that the tensorial
structure of the hydrodynamic interactions plays no important role
in the dynamics.

To analyze the motion of the system during one complete period of
the non-reciprocal cycle, we introduce an auxiliary move as shown
in Fig. \ref{auxiliary}. During this motion, the right arm has a
constant length $\delta$ while the left arm changes its length
from $D$ to $D-\epsilon$ with the constant velocity $W$. Using
symmetry arguments, we can relate all the four steps in the
non-reciprocal cycle to the above move, as follows: The step (a)
corresponds to the auxiliary move by setting $\delta=D$. By
applying a reflection transformation with respect to any point on
the $x$ axis, the step (b) can be mapped onto the auxiliary move
with $\delta=D-\epsilon$. The step (c) is obtained by applying a
time-reversal transformation on the auxiliary move, with
$\delta=D-\epsilon$.  Finally, the step (d) is obtained by
applying a reflection transformation [as in step (b)] followed by
a time-reversal transformation, with $\delta =D$.

To obtain the net displacement of the middle sphere in the real
problem, it is thus enough to solve the dynamical equation for a
single auxiliary movement. If we define the net displacement of
the middle sphere during the auxiliary step by $\Delta_{\rm
f}(\delta)$, then by considering the above arguments we can
calculate the total displacement $\Delta$ of the real system
through a complete cycle as
\begin{equation}
\Delta=2[\Delta_{\rm f}(D)-\Delta_{\rm
f}(D-\epsilon)].\label{Delta}
\end{equation}
Noting that the above displacement takes place during the time of
a complete cycle that is $4 \epsilon/W$, then we can calculate the
average swimming velocity for the swimmer as
\begin{equation}
V_{s}=W \; \frac{\Delta}{4\epsilon}.\label{Vs1}
\end{equation}

\begin{figure}
\includegraphics[width=.7\columnwidth]{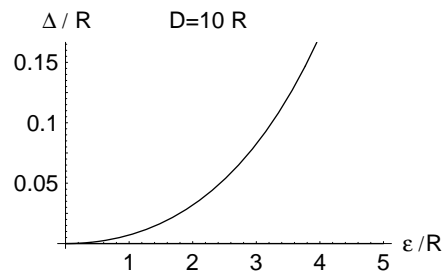}
\caption{Dimensionless displacement of the swimmer in a complete
cycle as a function of the dimensionless relative displacement
between neighboring spheres.} \label{graph}
\end{figure}

As an example, we have numerically calculated the displacements
for the case of $D=10 \; R$ and $\epsilon=4 \; R$. During the
first stroke the middle sphere swims in the $-x$ direction by an
amount $1.35 \;R$, while in the second stroke it swims a distance
of $1.44 \; R$ in the positive direction. The third stroke then
causes a continuation of the motion in the positive direction for
a distance of $1.44 \; R$, and finally, the fourth stroke takes it
back by a distance of $1.35 \;R$. At the end of this cycle the
sphere is displaced by a net amount of $0.16 \; R$ in the positive
direction, as shown in Fig. \ref{nonrecipcycle}. Figure
\ref{graph} shows the total displacement of the middle sphere
during a complete non-reciprocal cycle as a function of the
internal relative displacement $\epsilon$ for $D=10 \;R$.

In the limit of small internal deformation of the swimmer, we can
expand all the quantities in terms of $\epsilon/D$, and calculate
the swimming velocity in a perturbative series. To the leading
order, we find
\begin{equation}
V_{s}=0.7 \; W \; \left(\frac{R}{D}\right) \;
\left(\frac{\epsilon}{D}\right)^{2}.\label{Vs2}
\end{equation}
The above result shows that the scale of the swimming velocity is
set by the typical velocity of the internal motion. Moreover, the
swimming appears to be a quadratic effect with respect to the
small deformations in the system. These two characteristics are
general, as can be seen in other swimmers at low Reynolds number.

In our simple prescription for the non-reciprocal motion we have
assumed that in each step one sphere moves with respect to the
middle one and the other one is kept in relative constant
distance. This assumption has been made for simplicity, and we can
imagine a more general continuous motion of the spheres with
respect to the middle one. The only necessary condition is to
break the time-reversal symmetry and obtain a non-reciprocal
motion. The general requirement in the continuous motion will be
the existence of a nonzero phase difference between the continuous
periodic motions of the left and right spheres with respect to the
middle one. The swimming velocity in this case will also have the
two general characteristics mentioned above.

The generalized case of many spheres that are coupled to each
others on a regular one- or two-dimensional lattice is a suitable
microscopic model for an extensible flat filament or membrane.
In these cases, the relative ``in-plane'' motion of the
neighboring spheres with respect to each other will cause the
whole system to swim. The direction of motion and the swimming
velocity can be predicted using the simple proposed description.
The internal relative motion of the system that can cause swimming
is of the general form of a traveling wave on the position of the
spheres in some direction. Traveling wave is the simplest
non-reciprocal motion which breaks the time-reversal and
translational symmetries. The swimming velocity will be opposite
to the direction of the wave-vector, and proportional to the phase
velocity of the traveling wave. Such kind of motion has been
considered for a spherical membrane, and shown to have the same
characteristics as suggested above \cite{howard2}.

In conclusion we have introduced a very simple swimmer and
calculated its swimming velocity. The swimmer uses some periodic
internal motion to propel itself under low Reynolds number
conditions. The advantage of this model, as compared to previously
known model swimmers, is that the analysis of the hydrodynamics
problem can be performed considerably more easily. The model
swimmer could be used in making molecular-size machines with
controllable motion.

We have benefited throughout this work from fruitful discussions
with A. Ajdari, F. Julicher, K. Kruse, and H. Stone.


\end{document}